\documentclass[preprint]{aastex}
\usepackage{psfig}
\begin{document}
\title{Discovery of a new, third kHz QPO in 4U~1608--52, 4U~1728--34,
and 4U~1636--53\\ Sidebands to the lower kHz QPO?}

\author{Peter G. Jonker\altaffilmark{1}, Mariano
M\'endez\altaffilmark{1,2}, Michiel van der Klis\altaffilmark{1}}

\altaffiltext{1}{Astronomical Institute ``Anton Pannekoek'',
University of Amsterdam, and Center for High-Energy Astrophysics,
Kruislaan 403, 1098 SJ Amsterdam; peterj@astro.uva.nl,
michiel@astro.uva.nl, mariano@astro.uva.nl} 
\altaffiltext{2}{Facultad de Ciencias Astron\'omicas y
Geof\'{\i}sicas, Universidad Nacional de La Plata, Paseo del Bosque
S/N, 1900 La Plata, Argentina}

\begin{abstract}
\noindent
We report the discovery of a third kilohertz quasi--periodic
oscillation (kHz QPO) in the power spectra of the low--mass X--ray
binaries 4U~1608--52 (6.3$\sigma$), 4U~1728--34 (6.0$\sigma$), and
4U~1636--53 (3.7$\sigma$) which is present simultaneously with the
previously--known kHz QPO pair. The new kHz QPO is found at a
frequency that is 52.8$\pm$0.9 Hz, 64$\pm$2 Hz, 58.4$\pm$1.9 Hz higher
than the frequency of the lower kHz QPO in 4U~1608--52, 4U~1728--34,
and 4U~1636--53, respectively. The difference between the frequency of
the new kHz QPO and the lower kHz QPO increased in 4U~1608--52 from
49.6$\pm$1.4 Hz to 53.9$\pm$0.5 Hz when the frequency of the lower kHz
QPO increased from 672 Hz to 806 Hz. Simultaneously the difference
between the frequency of the new kHz QPO and the upper kHz QPO
increased by $\sim$60 Hz, suggesting that the new kHz QPO is unrelated
to the upper kHz QPO. In 4U~1636--53 a fourth, weaker, kHz QPO is
simultaneously detected (3$\sigma$) at the same frequency separation
below the lower kHz QPO, suggesting the new kHz QPOs are sidebands to
the lower kHz QPO.  We discuss the nature of this new kHz QPO and its
implications on the models for the kHz QPOs.
\end{abstract}


\section{Introduction}
\label{intro}
\noindent
In the last four years observations with the {\it Rossi X--ray Timing
Explorer} (RXTE) satellite have revealed the presence of several
quasi--periodic phenomena at frequencies higher than 100 Hz in the
Fourier power spectra of low--mass X--ray binaries (LMXBs). First, the
kilohertz quasi--periodic oscillations (kHz QPOs) were discovered (van
der Klis et al. 1996a,b; Strohmayer, Zhang, \& Swank 1996; Strohmayer
et al. 1996; for the most recent review see van der Klis
2000). Somewhat later, nearly coherent oscillations were discovered
during type I X--ray bursts in several of these LMXBs (Strohmayer et
al. 1996; for a review see Swank 2000); these burst oscillations
presumably occur at frequencies close to the neutron star spin
frequency (Strohmayer et al. 1996).  \par
\noindent
The kHz QPOs are nearly always found in pairs. Although their
frequencies can vary over several hundred Hz, the frequency separation
between the twin kHz QPOs, $\nu_2-\nu_1$, remains approximately
constant, close to the inferred spin frequency of the neutron star
(but see e.g. van der Klis et al. 1997; M\'endez et al. 1998a;
M\'endez, van der Klis, \& van Paradijs 1998; M\'endez \& van der Klis
1999).  Various models exist for the origin of these QPOs. Immediately
after their discovery a beat frequency model was proposed (Strohmayer
et al. 1996), of which the sonic--point model is the most elaborate
example (Miller, Lamb, \& Psaltis 1998). Later, Stella \& Vietri
(1998) proposed the relativistic precession model (see also Psaltis \&
Norman 2000; but see Markovi\'c \& Lamb 2000) and Osherovich \&
Titarchuk (1999) introduced the two--oscillator model. In the latter
model the QPO at $\nu_1$ (the lower kHz QPO) occurs at the Keplerian
frequency of material orbiting the neutron star, whereas in the other
two models the QPO at $\nu_2$ (the upper kHz QPO) is the one that is
Keplerian.  \par
\noindent 
The possibility that sidebands to the two main kHz peaks could occur
was mentioned by several authors (Miller et al. 1998; Psaltis \&
Norman 2000; Miller 2000), but no sidebands, nor in fact any other kHz
QPOs beyond the initial pair were detected up to now (see e.g.,
M\'endez \& van der Klis 2000).  In the LMXBs 4U~1608--52,
4U~1728--34, and 4U~1636--53 the kHz QPOs have been extensively
studied on previous occasions (4U~1608--52, Berger et al. 1996; Yu et
al. 1997; M\'endez et al. 1998a,b; 4U~1728--34, Strohmayer et
al. 1996; Strohmayer, Zhang, \& Swank 1996; Ford \& van der Klis 1998;
M\'endez \& van der Klis 1999; 4U~1636--53, Zhang et al. 1996;
Wijnands et al. 1997; M\'endez, van der Klis, \& van Paradijs 1998;
Kaaret et al. 1999, M\'endez 2000).
\newline
\noindent
In this Letter, we describe the discovery of a new, third kHz QPO
close to the lower kHz QPO in these three sources, which is probably
an upper sideband to the lower kHz QPO. We discuss briefly the
possible implications of this discovery in terms of the existing
models for the kHz QPOs.

\section{Observations and analysis}
\label{analysis}
\noindent
We have used observations obtained with the proportional counter array
(PCA; Jahoda et al. 1996) onboard the RXTE satellite (Bradt,
Rothschild, \& Swank 1993). The observations of 4U~1608--52 used here
are the same as those described by M\'endez et al. (1999, and
references therein); the observations of 4U~1728--34 are those used by
Strohmayer et al. (1996), Strohmayer, Zhang, \& Swank (1996), Ford \&
van der Klis (1998), and M\'endez \& van der Klis (1999). The
observations of 4U~1636--53 used in this paper are listed in
Table~\ref{obs_log}.  \par\noindent We used data with a time
resolution of at least 122$\mu$s to calculate power spectra of data
segments of 64 s up to a Nyquist frequency of 4096 Hz in two different
energy bands, from 2.0--8.7 keV and from 8.7--60 keV if available, and
in one band combining the total effective PCA energy range (2.0--60
keV). Observation 10094-01-01-00 of 4U~1608--52 only covered the
2.0--13.2 keV band and we excluded it in the calculation of the
fractional rms amplitudes of QPOs.  We selected only those power
spectra in which a narrow (full--width at half maximum, FWHM, less
than $\sim$10 Hz) lower kHz QPO was observed. This resulted in
$\sim$67, $\sim$92, and $\sim$80 ksec of data in the 2.0--60 keV range
which corresponds to $\sim$18\%, $\sim$12\%, and $\sim$15\% of the
total amount of analyzed data for 4U~1608--52, 4U~1728--34, and
4U~1636--53, respectively. 

\begin{deluxetable}{ll}
\tablecaption{Log of the observations of 4U~1636--53 used in this
analysis. For the observations we used in our analysis of the other
sources see main text.\label{obs_log}}
\tabletypesize{\normalsize}
\tablecolumns{2}
\tablewidth{0pc}
\tablehead{
\colhead{Observation} & \colhead{Date} \&  \\
\colhead{ID} & \colhead{Start time (UTC)}}
\startdata               
10088-01-01-00  & 	27-04-1996	 13:45 \\
10088-01-02-00  & 	29-04-1996	 17:37 \\
10088-01-03-00  & 	30-04-1996	 16:03 \\
10088-01-07-01  & 	09-11-1996	 20:53 \\
10088-01-08-03  & 	31-12-1996	 18:23 \\
10088-01-06-01  &  06-01-1997	 05:58 \\
10088-01-06-07  & 	06-01-1997	 08:32 \\
30053-02-01-000 & 	24-02-1998	 23:26 \\
30053-02-01-001 & 25-02-1998	 05:55 \\
30053-02-02-02  & 19-08-1998	 08:15 \\
30053-02-02-01  & 19-08-1998	 13:03 \\
30053-02-01-01  & 20-08-1998	 01:50 \\
30053-02-01-02  & 20-08-1998	 03:26 \\
\enddata
\end{deluxetable}

\par
\noindent 
We traced the lower kHz QPO using a dynamical power spectrum (e.g. see
plate 1 in Berger et al. 1996) displaying consecutive power spectra to
visualize the time evolution of the QPO frequency. For each source
separately, we fitted this QPO peak in each power spectrum in the
range 100 Hz above and 100 Hz below the traced QPO frequency (so to
exclude the upper kHz QPO) with a constant plus a Lorentzian. For each
source separately, we used the shift-and-add method described by
M\'endez et al. (1998b) to shift each lower kHz QPO peak to the same
frequency and average the aligned power spectra.  We then fitted each
average power spectrum (one per source) with a constant plus
Lorentzians to represent the QPOs. Errors on the fit parameters were
calculated using $\Delta\chi^2=1.0$ (1$\sigma$ single parameter). The
95\% confidence upper limits were determined using
$\Delta\chi^2=2.71$.

\section{Results}
\label{result}
\begin{figure*}
\centerline{\psfig{file=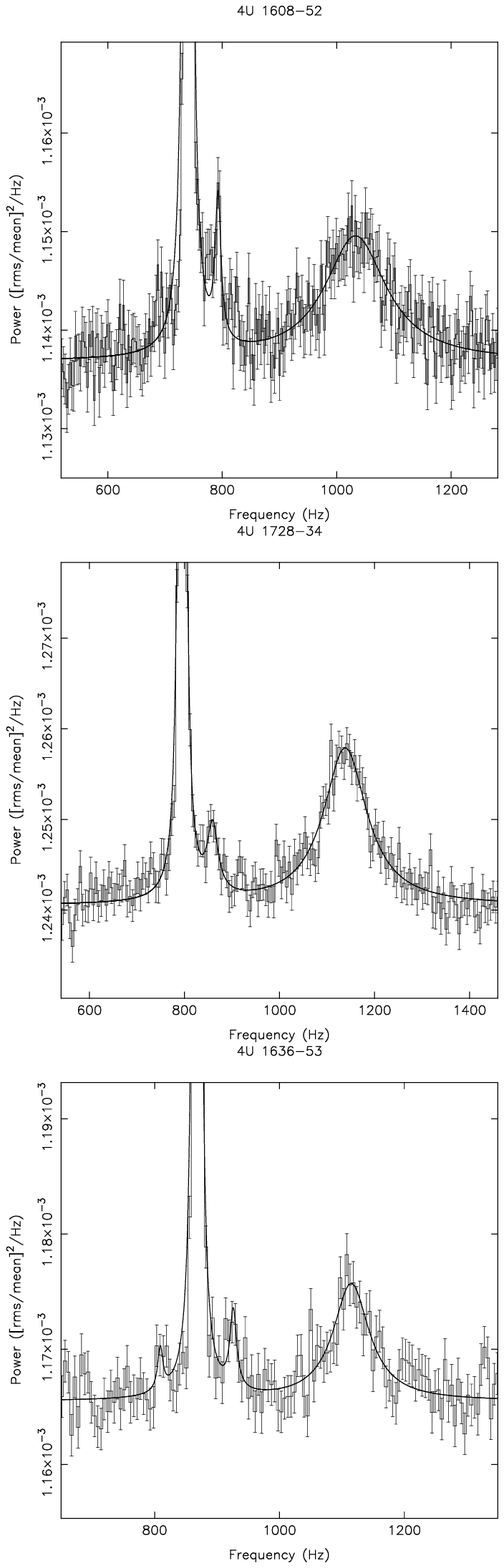,width=7cm}}\figcaption{Power density
spectra of the three sources showing the lower and upper kHz QPO and
to the high frequency side of the lower kHz QPO the new QPO. The power
is rms normalized after Belloni \& Hasinger (1990). Note that due to
the applied shift--and--add method, only the frequency difference with
respect to the frequency to which the power spectra were shifted is
meaningful (see also Table~\ref{prop_QPOs}).
\label{sidebands}}
\end{figure*}

\noindent
Figure~\ref{sidebands} shows the resulting power spectra for
4U~1608--52, 4U~1728--34, and 4U~1636--53 after applying the
abovementioned procedure.  We discovered a new kHz QPO in these three
sources at a frequency that is, respectively, 52.8$\pm$0.9 Hz,
64$\pm$2 Hz, and 58.4$\pm$1.9 Hz higher than that of the lower kHz
QPO, at a significance level of 6.3$\sigma$, 6.0$\sigma$, and
3.7$\sigma$ (single trial), respectively (Table
~\ref{prop_QPOs}). This new kHz QPO is detected at the same time as
the twin kHz QPOs that were already known in these sources.
\par
\noindent
In 4U~1636--53 we detected a fourth kHz QPO, at the same frequency
separation below the lower kHz QPO (Figure~\ref{sidebands}, Table
~\ref{prop_QPOs}). This new kHz QPO was detected in a single trial, at
a 3$\sigma$ significance level. The presence of two, symmetrically
located, peaks on either side of the lower kHz QPO suggests the new
kHz QPOs are sidebands to the main peak and from now on we will refer
to these new kHz QPOs as such. No significant lower sidebands could be
detected in the other two sources. 
\par
\noindent
We applied an F--test to the $\chi^2$ of the fits with and without the
upper sideband in order to test its significance.  We derived values
for the significance of the upper sideband similar to those calculated
from the errors in the fit parameters which we quoted
above. Conservatively estimating the number of trials involved in
obtaining our results at $\sim$400 (the number of frequencies where we
searched for a QPO, $\sim$4000, divided by the FWHM of the QPO
$\sim$10 Hz) still results in a $>5\sigma$ detection in case of
4U~1608--52.  \par
\noindent
For each source we divided the data into two parts based on the
frequency of the lower kHz QPO in the power spectra. In each of these
parts we shifted the lower kHz QPO peak to the same frequency as
before and fitted the average power spectrum. In 4U~1636--53 and
4U~1728--34 the separation between the upper sideband and the main
peak, $\Delta_{SB}$, is consistent with being the same in the two
parts, whereas in 4U~1608--52 $\Delta_{SB}$ increases significantly
(2.9$\sigma$, Table ~\ref{prop_QPOs}). In 4U~1608--52, the frequency of
the upper sideband increases by $\sim$139 Hz as the frequency of the
lower kHz QPO increases by $\sim$134 Hz. The fact that the distance of
the sideband remains approximately constant (to within $\sim$4.5 Hz)
as both peaks move over $\sim$135 Hz is an additional argument
supporting the interpretation of the new kHz QPO as a sideband to the
lower kHz QPO. Simultaneously, the frequency distance of the sideband
to the upper kHz QPO changes by nearly 60 Hz; this and the fact that
the sideband is much narrower than the upper kHz QPO, strongly argues
in favor of the sideband being un--related to the upper kHz QPO.  \par
\noindent
To correct for the relative motion of the sideband with respect to the
main peak, which artificially broadened it in our previous analysis,
we calculated the frequencies to which each power spectrum should be
shifted in order to align the sideband, assuming a linear relation
between the frequencies of the sideband and the main peak. Applying
this new shift to the data of 4U~1608--52 we indeed obtained a
narrower sideband (9.7$\pm$2.2 Hz), although the effect is
marginal. Aligning the power spectra on a hypothetical lower sideband
did not result in a significant detection in 4U~1608--52.  The upper
sideband is also significantly detected in 4U~1608--52 when only a
fraction of the data is selected and no shift-and-add is applied,
although at a smaller significance level.
\par
\noindent
Applying the same shifts to the power spectra calculated for the two
energy bands, we measured the dependence of the fractional rms
amplitude on photon energy of the detected QPOs. Only for 4U~1608--52
we were able to significantly detect the sideband in both energy
bands. The increase in fractional rms amplitude with energy in
4U~1608--52 is different for the lower and upper kHz QPO (see M\'endez
et al. 1998b).  Although the fractional rms amplitude of the sideband
increased from 1.7\%$\pm$0.2\% to 2.6\%$\pm$0.3\% as a function of
photon energy (2.0--8.7--20 kev) in 4U~1608--52, the uncertainties on
the measurements of the fractional rms amplitudes of the sideband did
not enable us to distinguish between the two different trends observed
for the kHz QPO pair. The FWHM of the lower kHz QPO, that of the upper
kHz QPO, and that of the sideband nor the sideband separation nor the
peak separation between the upper and lower kHz QPO changed
significantly with respect to the values obtained in the total energy
band.\par
\noindent
We also investigated the lower frequency part of the power spectra for
the presence of a QPO at a frequency equal to the difference between
the sideband and the lower kHz QPO. We detected a low frequency QPO
(LFQPO) at 40.9$\pm$0.7 Hz with a FWHM of 19$\pm$2 Hz and a fractional
rms amplitude of 2.8$\pm$0.1\% in 4U~1608--52 (see also Psaltis,
Belloni, \& van der Klis 1999). In 4U~1728--34, a LFQPO at
41.5$\pm$0.2 Hz, with a FWHM of 19.9$\pm$0.8 Hz and a fractional rms
amplitude of 5.3$\pm$0.1\% was present (see also Ford et al. 1998; Di
Salvo et al. 2000). In 4U~1636--53 no LFQPO was found, but a broad
noise component was detected which could be described with an
exponentially cutoff power law with a cut off frequency of 70$\pm$18
Hz, a power law index of 0.0$\pm$0.1, and a fractional rms amplitude
of 2.8$\pm$0.1\%.

\begin{deluxetable}{llllllllllll}
\tabletypesize{\tiny}
\tablecolumns{12}
\tablewidth{0pc}
\tablecaption{The properties (2.0--60 keV) of the lower ($\nu_1$),
upper ($\nu_2$) kHz QPO peaks, and of the new kHz QPO ($\nu_{SB}$) for
4U~1608--52, 4U~1728--34, and 4U~1636--53.\label{prop_QPOs}}
\tablehead{
\colhead{}  & \multicolumn{3}{c}{4U~1608--52} & \colhead{} &
\multicolumn{3}{c}{4U~1728--34} & \colhead{} &
\multicolumn{3}{c}{4U~1636--53} \\
\cline{2-4} \cline{6-8} \cline{10-12} \\
\colhead{Parameter} &\colhead{$\sim$67 ksec$^d$} &\colhead{$\sim$33
ksec$^e$} &\colhead{$\sim$34 ksec$^e$} &\colhead{} &\colhead{$\sim$92 ksec$^d$}
&\colhead{$\sim$42 ksec$^e$} &\colhead{$\sim$50 ksec$^e$} &\colhead{}
&\colhead{$\sim$80 ksec$^d$} &\colhead{$\sim$40 ksec$^e$}
&\colhead{$\sim$39 ksec$^e$} }  
\startdata
rms$_1$ (\%)& 8.47$\pm$0.02 & 8.66$\pm$0.03 & 8.89$\pm$0.02 & & 
6.71$\pm$0.02 & 6.56$\pm$0.04 & 6.99$\pm$0.03 & &  6.66$\pm$0.02 &
7.30$\pm$0.02 & 5.98$\pm$0.03 \\
FWHM$_1$ (Hz)	& 4.85$\pm$0.03	& 5.45$\pm$0.06 &4.38$\pm$0.04 & & 
7.5$\pm$0.1 & 9.1$\pm$0.1 & 6.6$\pm$0.1 & & 4.75$\pm$0.04 &
4.28$\pm$0.04& 5.8$\pm$0.1 \\
$\nu_1$ (Hz)$^a$ & 740 & 672 & 806 & & 795 & 733 & 847 & & 867.5 & 843 & 893 \\
rms$_2$ (\%)& 5.1$\pm$0.2 & 6.2$\pm$0.2 & 4.0$\pm$0.2 & & 
5.5$\pm$0.1 & 6.1$\pm$0.1 & 4.9$\pm$0.3 & & 3.3$\pm$0.2 & 3.7$\pm$0.2&
3.1$\pm$0.2 \\
FWHM$_2$ (Hz)	& 131$\pm$10 & 105$\pm$8 & 130$\pm$18 & &
111$\pm$6 & 95$\pm$5 & 141$\pm$20 & & 70$\pm$7 & 99$\pm$15 & 51$\pm$9 \\
$\Delta\nu$ (Hz)$^b$ & 293$\pm$3 & 308$\pm$3 & 253$\pm$6 & & 
343$\pm$2 & 350$\pm$2 & 324$\pm$5 & & 247$\pm$3 & 260$\pm$5 & 242$\pm$3 \\
rms$_{SB}$ (\%)& 1.77$\pm$0.14 & 2.0$\pm$0.2 & 1.3$\pm$0.1 & &
1.79$\pm$0.15 & 2.1$\pm$0.2 & 1.4$\pm$0.2 & & 1.18$\pm$0.17 & 1.1$\pm$0.2 & 1.5$\pm$0.2\\
FWHM$_{SB}$ (Hz) & 12$\pm$3 & 15$\pm$4 & 5$\pm$2 & & 
27$\pm$7 & 38$\pm$10 & 15$\pm$6 & & 13$\pm$6 & 10$\pm$3 & 27$\pm$10 \\
$\Delta_{SB}$ (Hz)$^c$ & 52.8$\pm$0.9 & 49.6$\pm$1.4 & 53.9$\pm$0.5 & &
64$\pm$2 & 65$\pm$4 & 65$\pm$2 & & 58.4$\pm$1.9 & 56.6$\pm$2.7 &
62.4$\pm$5.1 \\
rms$_{SB2}$ (\%)$^f$ & $<$0.9 & $<$1.2 & $<$1.1 & & $<$1.3 & $<$1.5 &
$<$0.9 & &
0.8$\pm$0.2 & 1.0$\pm$0.1 & $<$1.0\\
\tablenotetext{a}{Average frequency for the lower kHz QPO before shifting.}
\tablenotetext{b}{Frequency separation, $\Delta\nu = \nu_2-\nu_1$,
between the upper and lower kHz QPO.}
\tablenotetext{c}{Frequency separation, $\Delta_{SB} =
\nu_{SB}-\nu_1$, between the upper sideband and the lower kHz QPO.}
\tablenotetext{d}{Measurements for all available data}
\tablenotetext{e}{Data were divided according to the frequency of the
lower kHz QPO.}
\tablenotetext{f}{95\% confidence upper limits to the amplitude of the
lower sideband were calculated by fixing the other parameters at the
values obtained for the upper sideband.}
\enddata
\end{deluxetable}

\section{Discussion}
\noindent
We have discovered a new kHz QPO at frequencies close to the frequency
of the lower kHz QPO in the three LMXBs 4U~1608--52, 4U~1728--34, and
4U~1636--53. This new kHz QPO moves in frequency with the lower kHz
QPO and maintains a distance to it which is nearly (but not exactly)
constant; its distance to the upper kHz QPO varies much more.  In
4U~1636--53 an additional QPO was found symmetrically located at the
lower frequency side of the lower kHz QPO peak (3$\sigma$). These
facts suggest that the new kHz QPOs are sidebands to the lower kHz
QPO.\par
\noindent
If these QPOs are sidebands due to an amplitude modulation of the
lower kHz QPO, there must be an additional mechanism reducing the
amplitude of the lower sideband or enhancing the upper sideband, since
in 4U~1608--52 and 4U~1728--34 the presence of a symmetric lower
sideband can be excluded. If the lower kHz QPO is a beaming
oscillation, a single-sideband (rotational) beat frequency scenario
could apply (e.g. Alpar \& Shaham 1985). However, this would lead to a
lower rather than an upper sideband, opposite to what is observed,
unless the modulating part of the disk is counter rotating.  When both
rotational modulation and amplitude modulation of the formation of the
lower kHz QPO produce sidebands, but of opposite phase, destructive
interference could in principle cause the amplitude of the lower
frequency sideband to be suppressed. Kommers et al. (1998) observed
sidebands to the pulses of the pulsar 4U~1626--67 at much lower
frequencies ($\sim$0.1Hz). There the lower sideband was stronger than
the upper one and they proposed a model using both amplitude
modulation and rotational modulation of the pulsar beam.  \par
\noindent
The frequency separation between the sideband and the main peak
($\sim$50--60 Hz), hereafter `sideband separation' is reminiscent of
typical frequencies of horizontal branch oscillations (HBOs) in Z
sources and low frequency QPOs (LFQPOs) in atoll sources. However, the
frequency of the LFQPO ($\sim$41 Hz) apparent in two of the three
sources is inconsistent with the observed sideband
separation. Furthermore, the increase in sideband separation with
increasing lower kHz QPO frequency we observed in 4U~1608--52 is less
than the increase in HBO frequency with lower kHz QPO frequency
usually seen (Psaltis et al. 1999).  So, the sidebands appear to be
unrelated to the LFQPO in these atoll sources. Mechanisms that might
explain these QPOs, such as the magnetospheric beat frequency model
(Alpar \& Shaham 1985; Lamb et al. 1985) or the Lense-Thirring
precession model (Stella \& Vietri 1998) could be used in explaining
the LFQPO and the sideband, however one model can not explain both
simultaneously unless the LFQPO and the sideband reflect frequencies
at different radii in the disk.  \par
\noindent
None of the kHz QPO models in the literature (e.g. the sonic point
model, Miller et al. 1998, Miller 2000, the relativistic precession
model, Stella \& Vietri 1998, and the two--oscillator model,
Osherovich \& Titarchuk 1999) in their present form predict the
presence of an upper sideband to the lower kHz QPO with a sideband
separation different from the LFQPO frequency.  \par
\noindent
One of the effects possibly occurring in the sonic point beat
frequency model (Miller et al. 1998, Miller 2000) could perhaps
explain the formation of the upper sideband we discovered. The density
enhancements along the spiral flow responsible for the generation of
the lower kHz QPO rotate around the neutron star $\sim$5--10 times
before reaching the surface (Miller et al. 1998). Therefore, the
orbital frequency of the enhancements is higher than the orbital
frequency at the sonic point, the beat frequency will end up at
slightly higher frequencies than the frequency of the lower kHz
QPO. Since the sideband has a FWHM of $\sim$10 Hz corresponding to at
least $\sim$100 cycles at a Keplerian frequency of 1000 Hz, the beat
frequency will end up at frequencies 50--100 Hz higher. The orbital
frequency of the enhancements and the change in the sideband
separation as a function of the frequency of the lower kHz QPO, as
observed in 4U~1608--52, will depend on the details of the physical
processes at work close to the neutron star.

\acknowledgments This work was supported in part by the Netherlands
Organization for Scientific Research (NWO) grant 614-51-002. This
research has made use of data obtained through the High Energy
Astrophysics Science Archive Research Center Online Service, provided
by the NASA/Goddard Space Flight Center. This work was supported by
NWO Spinoza grant 08-0 to E.P.J.van den Heuvel. MM is a fellow of the
Consejo Nacional de Investigaciones Cient\'{\i}ficas y T\'ecnicas de
la Rep\'ublica Argentina. PGJ would like to thank Jeroen Homan for
stimulating discussions and Rob Fender for comments on an earlier
version of this work. We would like to thank the anonymous referee for
his/her comments which improved the paper.

\end{document}